\documentclass[letter]{aa} % for the letters 

\usepackage{graphicx}
\usepackage{txfonts}
\begin{document}

   \title{Dwarfs walking in a row.}

   \subtitle{The filamentary nature of the NGC~3109 association.}

   \author{M. Bellazzini\inst{1}, T. Oosterloo \inst{2,3}, F. Fraternali\inst{4,3},
          \and
          G. Beccari\inst{5}
          }

   \institute{INAF - Osservatorio Astronomico di Bologna, via Ranzani 1, 40127 Bologna, Italy\\
              \email{michele.bellazzini@oabo.inaf.it}
              \and
              Netherlands Institute for Radio Astronomy, Postbus 2, 7990 AA Dwingeloo, 
              The Netherlands
              \and
              Kapteyn Astronomical Institute, University of Groningen, Postbus 800, 9700 AV
              Groningen, The Netherlands
              \and
              Dipartimento di Fisica e Astronomia - Universit\`a degli Studi di Bologna, 
              viale Berti    Pichat 6/2, 40127 Bologna, Italy
              \and
             European Southern Observatory, Karl-Schwarzschild-Str. 2, 85748 Garching bei Munchen,
             Germany
             }

   \date{Accepted for publication by A\&A. Submitted on September 24, 2013.}

% \abstract{}{}{}{}{} 
% 5 {} token are mandatory
 
  \abstract{We re-consider the association of dwarf galaxies around NGC~3109,
whose known members were NGC~3109, Antlia, Sextans~A, and Sextans~B, based on a new updated
list of nearby galaxies and the most recent data. We find that the original
members of the NGC~3109 association, together with the
recently discovered and adjacent dwarf irregular Leo~P, form a very tight and elongated configuration in space. All these galaxies lie within $\sim 100$~kpc of a line that is
$\simeq 1070$~kpc long, from one extreme (NGC 3109) to the other (Leo~P), and 
they show a gradient in the Local Group standard of rest velocity with a total amplitude of
43~km~s$^{-1}$~Mpc$^{-1}$, and a r.m.s. scatter of just 16.8~km~s$^{-1}$. It is shown that the
reported configuration is exceptional given the known dwarf galaxies in the Local Group and
its surroundings.  We conclude that (a) Leo P is very likely an additional member of the 
NGC~3109 association, and (b) the association is highly ordered in space and velocity, and it is very elongated, suggesting that it was created by a tidal interaction or it was  accreted as a filamentary substructure.}


   \maketitle
%
%________________________________________________________________

\section{Introduction}

 The identification of planar alignments of dwarf satellites (and globular clusters) of the Milky Way (MW) and M31 has a long history, starting from \citet{LB} and  \citet{FFP}, respectively; since \citet{kro05} it has become one of the most lively research fields in Galactic astronomy. Indeed, very significant large-scale structures have recently been recognized within the Local Group (LG). For example, \citet{and_plane} discovered that a large fraction of M31 dwarf satellites lie in a vast, thin planar structure, that shows
large-scale rotation. \citet[][ hereafter P13]{kroupa}, found that in addition to the already known thin planes of dwarf galaxies surrounding both M31 and the MW, most of the LG non-satellite dwarfs are also located in two very large planar structures (diameters 1-2~Mpc). Accretion along cold filaments \citep{dekel} has been proposed as a possible origin for these ordered structures \citep[see, e.g.,][]{lovell} within the generally accepted $\Lambda$-cold dark
matter ($\Lambda$-CDM) paradigm for galaxy formation \citep[][]{bark}.
However several studies have reported that the $\Lambda$-CDM predictions are not compatible with the characteristics of the observed structures \citep[see, e.g.,][and references therein]{paw12}, proposing as an alternative that the planar alignments are the relics of huge tidal streams that arose from a major galaxy interaction that occurred in the LG \citep[][P13]{hammer}.

In this context it seems very important to recall the discovery by \citet[][hereafter T06]{tully06} that a significant fraction of dwarf galaxies in the local volume belong to associations of dwarfs, i.e., small
groups with strong correlation in space and velocity. T06 conclude that these
associations are probably not in dynamical equilibrium, but they can be
gravitationally bound, and, in this case, they are strongly dominated by dark matter, with
mass-to-light ratios in the range $\sim 100-1000~M_{\sun}L_{\sun}^{-1}$. Accretion of
 and/or interactions among dwarfs in associations may play
a role in the formation of coherent structures around giant galaxies
(\citealt{sales,dong08}; but see also \citealt{metz}).

The first bona-fide association discussed by T06 is the so called {\em NGC~3109
group}, from the gas-rich dwarf disk galaxy that is its most luminous member 
\citep[$M_V=-14.9$, according to][hereafter M12]{mcc}. The group was previously  
discussed in detail by \citet{vdb99}. It lies just beyond the zero
velocity surface of the LG, near the threshold between systems bound to the LG
and the local Hubble flow (see Fig.~5 in M12), in a quite isolated location 
(more than 1~Mpc apart from both the MW and M31; see Fig.~\ref{gen}, below). 
In addition to NGC~3109, the
other members of the association, recognized by T06 as lying at similar distance
and with similar radial velocity in the Local Group standard of rest ($V_{LG}$),
are the dwarf irregulars (dIrr) Sextans~A, Sextans~B, and Antlia, having
$M_V=-14.3,~-14.5$, and $-10.4$, respectively. T06 conclude that {\em ``...the
NGC~3109 association is the nearest distinct structure of multiple galaxies to
the Local Group.''} Recently \citet[][hereafter ST13]{shaya} noted that, since
all the members lie at very similar distances from us and in the same portion of
the sky (within $\sim 30\degr$ from NGC~3109) they effectively lie in a
common plane, nearly perpendicular to the line of sight. It is interesting
to note that none of the group members has been associated with the vast planar
structures discussed by P13. 

Triggered by the discovery of possible tidal tails in the outskirts of the
group member Sex~A (Bellazzini et al., in preparation), we reconsidered the
NGC~3109 association in the light of the most recent data, summarized in the
thorough catalogue of LG and nearby galaxies by M12 
(in particular, all the known galaxies within 3~Mpc of the MW). Interactions within the
group may be responsible for tidal features in some member \citep[see,
e.g.,][]{dong09,penny}. We suddenly realized that the newly discovered faint dIrr
Leo~P \citep{giova,rhode} is a likely member of the group\footnote{Thus the
discovery of Leo~P possibly vindicates the last sentence of \citet{vdb99}, although the Leo
constellation was not explicitly mentioned: {\em ``...It would be of interest to
search for additional faint members of the Ant-Sex group in Antlia, Hydra, and
Sextans.''}} and that all the known members, including Leo~P, appear to be
strictly clustered {\em along a line in space}. This finding strongly suggested
the opportunity of a more systematical revision of the association: the results
of this analysis are the subject of the present letter. It must be stressed that, in the following analysis, lines are adopted just as simple and convenient tools to parametrize and quantify the tightness of configurations that appear to be very elongated in one direction in space.

\section{Analysis}

All the data considered in the following analysis are drawn from the M12
compilation, except for (a) the distances to NGC~3109, Sex~A, Sex~B, and Antlia, and the associated uncertainties, that are from the homogeneous set by \citet{dal09}, and (b) the distance to Leo~P that is still highly uncertain, since available estimates range from 1.3~Mpc \citep{giova} to
1.5-2~Mpc \citep{rhode}. Here we adopted $D=1.5\pm0.5$~Mpc, for
simplicity. 

In the following, we will adopt a Cartesian reference frame
($X_a,Y_a,Z_a$) centered on NGC~3109, obtained by translating the origin of
the classical Galacto-centric reference frame 
\citep[as defined in Eq. 3 and 4 by][adopting $R_{\sun}=8.0$~kpc]{thomas} to the position of NGC~3109. We also define

$R= \left\{
\begin{array}{l@{\quad}l}
\sqrt{X_a^2+Y_a^2+Z_a^2} & Y_a \ge 0\\~\\
-\sqrt{X_a^2+Y_a^2+Z_a^2} & Y_a < 0
\end{array}\right.$

In case of aligned configurations, this is essentially a distance {\em along the line} in space, 
providing a more fundamental and general coordinate than $X_a,Y_a,Z_a$. For example, a velocity gradient along one of these coordinates would be just {\em a projection} of 
a  correlation between velocity and spatial distance along the line.
In the following we will always refer to velocity gradients (and corresponding rms velocity scatter about a best-fit line, rms$_V$) along this R coordinate, if not otherwise stated.

\begin{figure}
   \centering
   \includegraphics[width=8.5cm]{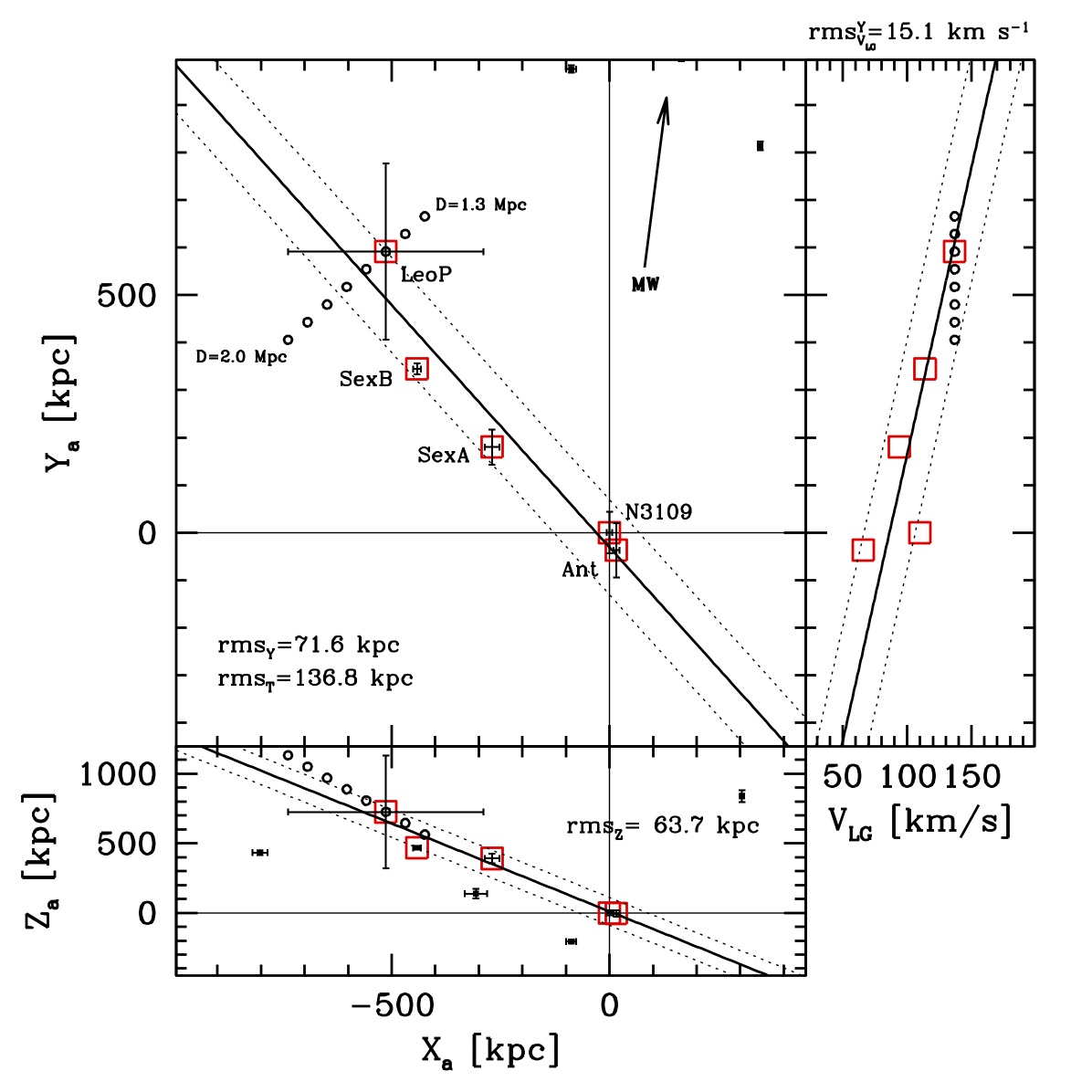}
      \caption{The alignment within the NGC~3109 association including Leo P.
The members of the association plus Leo~P are plotted as large empty squares; 
other galaxies from the M12 catalog (more distant than
300 kpc from the MW or M31) are plotted as small filled squares in the left
panels. Open circles marks the position of Leo~P with different
assumed values for its distance, from 1300~kpc to 2000~kpc, with steps of 100~kpc.
Thick lines are linear fits to the data in
each plane. The thin dotted lines enclose
the $\pm 100$~kpc range about the best-fit lines (left panels) or the $\pm
20$~km~s$^{-1}$ range about the best-fit line (right panel). The arrow indicate
the direction towards the MW.} 
%The rms about the best fit lines is reported in each panel.}
         \label{vLG}
   \end{figure}

In the left panels of Fig.~\ref{vLG} we plot the positions of the galaxies in the M12 catalog that lie more than 300~kpc away from M31 or the MW (to avoid unnecessary confusion) in the principal planes of the NGC~3109-centric reference frame
defined above. It is clear that NGC~3109+Antlia, Sex~A, Sex~B, {\em and} 
Leo~P are strictly aligned in both planes, {\em implying that
they are strongly clustered along a line in space}. The thick lines are linear
fits to these five galaxies. For simplicity, in all the considered cases we perform independent fits in each plane ($X_a$ vs. $Y_a$, and  $X_a$ vs. $Z_a$). Each linear fit defines a plane in space; the intersection of the two planes is a line in 3-D space. In this context we adopt the squared sum of the residuals of the two linear fits (rms$_T=\sqrt{\rm rms_Y^2+rms_Z^2}$) as a measure of the
global rms scatter about the line in space.
The maximum distance from the line is 98~kpc in $Y_a$, 101~kpc in $Z_a$, and rms$_T$ is 95.8~kpc,  while the length of the structure is $L=1067$~kpc, thus implying a 
length-to-thickness ratio $L/{\rm rms}_T=11.1$.

The right panel of Fig.~\ref{vLG} shows that the five 
galaxies also display a very tight correlation between position along the line
and $V_{LG}$, in the sense that galaxies farther from NGC 3109 have
larger recession velocities with respect to the LG. The total amplitude of the
gradient, as measured from the best-fit line in the R vs. $V_{LG}$ plane, is 45~km~s$^{-1}$ from one extreme to the other. This corresponds to 43~~km~s$^{-1}$Mpc$^{-1}$, not
compatible, at face value, with pure Hubble flow 
\citep[$H_0=67.3\pm 1.2$ km~s$^{-1}$Mpc$^{-1}$, see, e.g.,][]{ade}. The rms of the residuals is just
16.8~km~s$^{-1}$ ($rms_V^Y=$15.1~km~s$^{-1}$ in the projection shown in Fig.~\ref{vLG}), to be compared with a total velocity dispersion of $\sigma_{V_{LG}}=26.3$~km~s$^{-1}$. 
The effects of varying the distance to Leo~P in the range 1.3~Mpc - 2.0~Mpc is also displayed in Fig.~\ref{vLG} (small open circles). Considering that the linear fits would adjust according to the adopted distance, it appears that a remarkable alignment is preserved over a large range of distances; the tightest alignment would be reached for $D\simeq 1700-1800$\footnote{While the first version of this paper was under the scrutiny of the Referee, a preprint was posted \citep{mcq} reporting a more accurate estimate of the distance to Leo~P, $D=1.72^{+0.14}_{-0.40}$~Mpc, thus providing further support to the tightness of the alignment described here, and also suggesting Leo~P as a likely member of the NGC~3109 group.}~kpc. 

While considering the significance of spatial
alignments it must be remembered that NGC~3109 and Antlia are probably interacting 
and they may be a bound couple \citep{vdb99,penny};
hence NGC~3109+Antlia should be considered a single
system, in this sense\footnote{In Fig.~\ref{vLG} we consider each galaxy as a separate system for completeness and simplicity. This
is a conservative approach since a bound pair will artificially inflate the 
velocity spread of a structure in analogy to what happens with
binaries in stellar systems \citep{ols}.}. \citet{vdb99} suggested that 
Sex~A and B may also have interacted in the past, but today they are more than
250~kpc apart and so we consider them to be individual galaxies.
For the statistical analysis performed in
Sect.~\ref{signi}, thus, it is worth reporting that rms$_T=110.2$~kpc and
$rms_V=13.8$~km~s$^{-1}$, if Antlia is excluded from the sample. The R vs. $V_{LG}$ correlation remains strong, while the amplitude of the gradient reduces to 25~~km~s$^{-1}$Mpc$^{-1}$.
None of the other galaxies in the left panels of Fig.~\ref{vLG} seems associated with the
described alignment. The latter also does not appear to have any obvious relation with the known planar structures of the LG (P13). For brevity, in the following we will refer to the NGC~3109 group {\em plus} Leo~P as the {\em extended NGC~3109 association}.

%
%                                                One column figure
%----------------------------------------------------------- mappLG
%
%______________________________________________________________

\subsection{Can the observed alignment arise by chance?}
\label{signi}

What is the statistical significance of the configuration described above? 
This is very hard to assess in general, and in particular in the present case, 
given the sizable errors that affect the distance estimate of Leo~P, and the
effects of incompleteness as a function of distance and luminosity
\citep{tolle}. 
In an attempt to overcome some of these issues we proceed in 
a fully non-parametric way. 
 
\begin{figure}
   \centering
   \includegraphics[width=7cm]{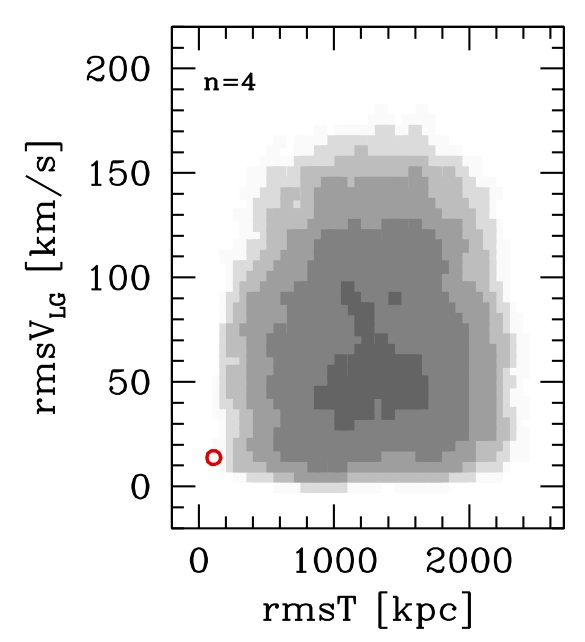}
      \caption{Density distribution of all the explored configurations of groups of four
      galaxies in the rms$_V$ vs. rms$_T$ space, compared to the position of
      the extended NGC~3109 association without Antlia (n=4).
      The density increases by a factor of two with each darker tone of gray. The contour of the lightest shade of gray includes more than 99.5 per cent of the cases.
             }
         \label{vet}
   \end{figure}

\begin{figure}
   \centering
   \includegraphics[width=8.0cm]{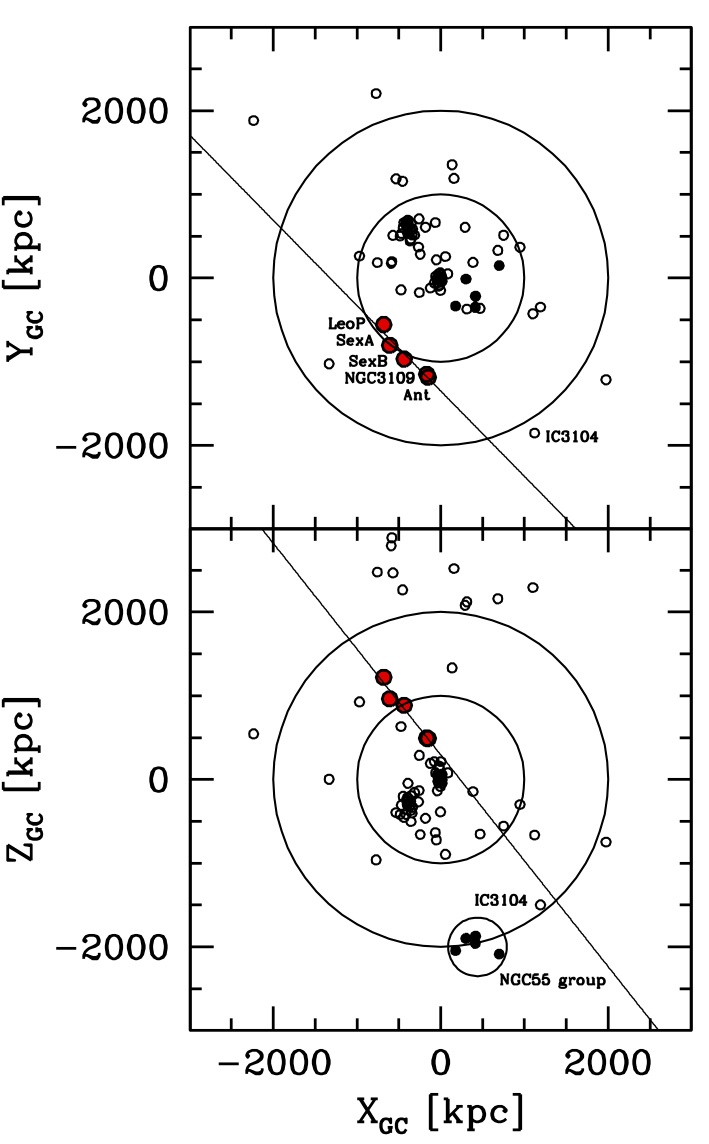}
      \caption{Galaxies in the M12 list plotted in the Galactocentric Cartesian reference frame
      (small open circles).
      The dwarfs of the {\em extended} NGC~3109 association are plotted as large red filled circles
      (with labels); those belonging to the NGC~55 association are plotted as dark filled circles.
      The symbol corresponding to IC~3104 has also been labeled.
      The lines are the best-fit lines to the members of the {\em extended} NGC~3109 association
      (including Antlia). The wide concentric circles have radii of 1~Mpc and 2~Mpc.}
         \label{gen}
   \end{figure}

Among the galaxies listed in the M12 catalog, we include in our analysis only those that have valid estimates of the distance and radial velocity, and that lie more than 300 kpc away from M31 or the MW, in order to avoid trivial alignments
among dwarfs strongly clustered in the respective subsystems. Antlia and 
HIZSS3B were also removed as possible members of binary systems \citep{begum}. Then we
considered all the independent 66405 groups of four galaxies that can be
extracted from the 37 galaxies remaining in the sample. For each group we
perform the same change of reference system adopted for the NGC~3109
association above (selecting one of the four galaxies at random as the origin of
the coordinates), we find the best fit lines in the same principal planes as in
Fig.~\ref{vLG}, and we compute rms$_T$ and rms$_V$ to check if values as low as
those observed for the extended NGC~3109 association (without Ant) are common, and how
frequently  they could arise in this set of Local Group galaxies\footnote{We note that the vast majority of the randomly extracted groups do not correspond to real associations, hence the adopted procedure effectively compares the tightness of the considered structures against chance alignments that can occur in a sample including all the relevant selection effects, since it is the {\em real} sample. It would be impossible to implement these selection effects in mock catalogs, since they are not completely known.}. 
It turned out that alignments having rms$_T\le
130$~kpc and rms$_V\le20.0$~km~s$^{-1}$ occur only four times in 66405 cases. In
addition to the case under scrutiny, there are two combinations of members of the NGC~55 subgroup 
(already recognized as a real association by T06; see also M12), i.e., NGC~55+NGC~300+ESO~294-G010 plus IC~5152 or IC~4662 (the only galaxy not classified as a member of the group).
While interesting, these alignments appear dominated by three galaxies (NGC~55, NGC~300, and ESO~294-G010) having similar $V_{LG}$ and enclosed within a radius of just $\simeq 320$~kpc (see Fig.~\ref{gen}), i.e., such a strict relation that they can be considered an individual triple system in this context. On the other hand, Sex~A and Sex~B are $\simeq 510$~kpc and $\simeq 730$~kpc from NGC~3109 (we recall that Ant is excluded from this analysis).
The fourth configuration satisfying the conditions above is composed by NGC~3109+SexA+SexB plus 
IC~3104 instead of~Leo P. We conclude that (a)  the NGC 3109 linear group is unique in the
surroundings of
the LG, as illustrated in Fig.~\ref{vet}; (b) the very few cases of comparable tightness are not random at all, as they correspond to known associations; and (c) by transitive property, IC~3104
should also be approximately aligned with the {\em extended} NGC~3109 association.

The extended NGC~3109 association {\em plus} IC~3104 has rms$_T$=136.8~kpc along its best-fit line, and the overall length of the structure is 2200~kpc, corresponding to a
length-to-thickness ratio $L=16.0$. While the line is
virtually the same as that of Fig.~\ref{vLG}, IC~3104 does not fit into the
velocity gradient shown by the members of the extended NGC~3109 association.
Moreover, Fig.~\ref{gen} shows that, while NGC~3109+Ant, Sex~B, Sex~A, and Leo~P are adjacent one to the other, IC~3104 is far away from all of them, on the other side of the LG with respect to M31 and the MW, at least in the $X_{GC}$ vs. $Y_{GC}$ direction; clearly it cannot be presently bound to the association, and it is unlikely to be physically associated. We are inclined to consider IC~3104 an individual chance match approximately along the direction of maximum elongation of a real structure, i.e., the filamentary {\em extended} NGC~3109 association.

%
%                                                One column figure
%----------------------------------------------------------- mappLG
%
%______________________________________________________________

%
%                                                One column figure
%----------------------------------------------------------- mappGSR
%
%______________________________________________________________

\section{Summary and Discussion}

In this letter we re-considered the spatial distribution and the kinematics of the NGC~3109 association of dwarf galaxies. The main results of our analysis can be summarized as follows:

   \begin{enumerate}
   
\item The distance and velocity of the recently discovered faint dIrr galaxy Leo~P 
are compatible with the hypothesis that Leo~P is an additional member of the well-known NGC~3109 association. According to Eq.~4 in \citet{tully02}, the mass of the extended association would be $M\simeq 3.2\times 10^{11}~M_{\sun}$ (assuming that it is bound), quite typical for this kind of group (T06). According to Eq.~1 in T06, the turnaround diameter (in spherical approximation) is $2r_{dw}^{1t}\simeq 980$~kpc, comparable to the total side-to-side linear extension of the association ($\simeq 1070$~kpc). 
      
\item We found that the members of the {\em extended} NGC~3109 association
are tightly aligned in space; the length of the configuration is $\simeq 11$ times the global rms scatter about the line. 
The aligned galaxies also display a tight correlation between the distance along the
best-fit line and their velocity in the LG standard of rest, with rms$_V$=16.8~km~s$^{-1}$. Even if a member of the association (e.g., Leo~P) is no longer bound to it, the alignment and the velocity gradient support the idea that they were all part of the same physical structure in the past. 

\item Exploring all the possible alignments of groups of four galaxies, among those listed in the M12 catalogue and not strictly associated with the MW or M31, we find that alignments as tight as those described above are extremely rare and it appears very unlikely that they can occur by chance (see also Fig.~\ref{vet}). 

\item A tidal interaction would be a quite natural explanation for the origin of the
structure, either considering
an encounter with the MW \citep[as envisaged by ST13; see also][]{zhao} that may have stretched the intragroup distances of the original association in the plane of the orbit and produced the velocity gradient, or
considering the members of the group to be {\em tidal} dwarfs \citep{duc}, from the fragmentation of a large-scale 
tidal arm formed during a major galaxy interaction in the LG \citep[see, e.g.,][P13]{hammer}.
In the framework of Newtonian dynamics, the high dark matter content of at least some of the members \citep[][]{carig} would militate against the latter hypothesis, that can be a viable solution within other gravitation theories \citep{mcg}.
Instead, if we assume that the members of the extended NGC~3109 association have recently left the Hubble flow and are currently falling into the LG for the very first time, the possibility that the observed linear structure could have formed within a  thin and cold cosmological
filament \citep{dekel,lovell} should be also considered.

\end{enumerate}

\begin{acknowledgements}
M.B and F.F. acknowledge the financial support from the PRIN MIUR 2010-2011 project ``The
Chemical and Dynamical Evolution of the Milky Way and Local Group Galaxies'',
prot. 2010LY5N2T.
\end{acknowledgements}

%-------------------------------------------------------------------

\end{document}